\documentstyle[aps,preprint,epsfig,tighten]{revtex}
\catcode`\"=\active \let"=\"  \let\3=\ss
\newcommand{\be}{\begin{eqnarray}}
\newcommand{\ee}{\end{eqnarray}}
\newcommand{\beq}{\begin{equation}}
\newcommand{\eeq}{\end{equation}}
\newcommand{\ba}{\begin{array}{l}}
\newcommand{\ea}{\end{array}}

\newcommand{\bb}{\begin{thebibliography}}
\newcommand{\eb}{\end{thebibliography}}
\newcommand{\ci}[1]{\cite{#1}}
\newcommand{\bi}[1]{\bibitem{#1}}

\newcommand{\ov}{\over}

\newcommand{\CA}{{\cal A}}

\newcommand{\CD}{{\cal D}}

\newcommand{\CF}{{\cal F}}

\newcommand{\CH}{{\cal H}}
\newcommand{\CI}{{\cal I}}
\newcommand{\CJ}{{\cal J}}

\newcommand{\CM}{{\cal M}}

\newcommand{\CP}{{\cal P}}

\newcommand{\CR}{{\cal R}}
\newcommand{\CS}{{\cal S}}
\newcommand{\CT}{{\cal T}}

\newcommand{\CV}{{\cal V}}
\newcommand{\CW}{{\cal W}}
\newcommand{\CX}{{\cal X}}

\newcommand{\ga}{\alpha}
\newcommand{\gb}{\beta}
\newcommand{\g}{\gamma}

\newcommand{\gz}{\zeta}
\newcommand{\gl}{\lambda}
\newcommand{\gt}{\theta}

\newcommand{\gr}{\rho}
\newcommand{\gs}{\sigma}

\newcommand{\go}{\omega}
\newcommand{\gG}{\Gamma}
\newcommand{\gD}{\Delta}
\newcommand{\gL}{\Lambda}
\newcommand{\gT}{\Theta}
\newcommand{\gS}{\Sigma}

\newcommand{\gO}{\Omega}
\newcommand{\gve}{\varepsilon}
\newcommand{\gvt}{\vartheta}

\newcommand{\gvp}{\varphi}
\newcommand{\abstitle}[1]{{\small {\bf #1}}}
\newcommand{\absauthor}[1]{\small {\bf #1}}

\begin{document}
 \begin{center}
 \abstitle{{RELATIVISTIC AND POLARIZATION
 PHENOMENA IN
 $NN\to d\pi$ PROCESSES }
 \footnote{encouraged and supported by Russian Foundation of
 Fundamental Research}}\\[3.0mm]
 \absauthor{{A.Yu.Illarionov , G.I.Lykasov
 \footnote{alexej@nu2.jinr.ru, lykasov@nu2.jinr.ru}}}
 \\[2.0mm]
 {\it { Joint Institute for Nuclear Research,\\
 141980 Dubna, Moscow Region, Russia}}
 \end{center}

 \begin{abstract}
  A detailed analysis of processes of the type $NN\to d\pi$ is presented
  taking into account the exchange graphs of a nucleon and a pion.
  A large sensitivity of polarization observables to the off-mass shell
  effects of nucleons inside the deuteron is shown. Some of these
  polarization characteristics can change the sign by including these
  effects. The influence of the inclusion of a $P$-wave in the deuteron wave
  function is studied, too. The comparison of the calculation results of
  all the observables with the experimental data on the reaction
  $pp\to d\pi^+$ is presented.

 \end{abstract}

 \vspace{1cm}

 \noindent


 \vspace{1cm}

 \section{\bf Introduction}

 As known, pion production in $NN$ collisions, in particular the
 channel $NN\to d\pi$, has been investigated by many theorists and
 experimentalists over the last decades. An earlier study of this reaction
 \ci{watson,rosen} and \ci{mandel} show that the excitation
 of the $\Delta$-isobar is a crucial ingredient for explaining the
 observed energy dependence of the cross section. A lot of papers are
 based on multichannel Schr\"odinger equations with separable or local
 potentials \ci{nisk,green}, \ci{koltun,rinat} and \ci{blank}. However,
 this study was performed within the nonrelativistic approach. Early
 attempts to develop the relativistic approach were made in
 \ci{yao,cha}, \ci{schiff,barry}. Both the pole graph, i.e. one-nucleon 
 exchange, and the rescattering graph presented below were calculated in this
 paper. As shown (see, for example, \ci{barry}), this diagram can result in a
 dominant contribution to the cross section of the discussed process.
 By the calculation of this one, some approximations, in particular the
 factorization of nuclear matrix elements, neglect of recoil etc.,
 were introduced which lead to an uncertainty of the final results.
 A more careful relativistic study of the reaction $pp\to d\pi^+$ was made
 in \ci{locher,loch1,loch2,loch3}. The pole and rescattering graphs were 
 shown to be insufficient to describe the experimental data; high order 
 rescattering contributions  should be taken into account. However, in this 
 approach there was no successful description of all the polarization 
 observables, especially the asymmetries $A_{y0}$, $iT_{11}$.
 Really, analyzing reactions of the type $NN\to d\pi$, there occurs a
 problem related to the off-mass shell effects of nucleons inside the
 deuteron. When the pion is absorbed by a two-nucleon pair or the deuteron,
 the pion energy is divided between two nucleons. So, for example, the
 relative momentum of the nucleon inside the deuteron increases at least
 by a value $\sim\sqrt{m\mu}=360 MeV$ if the rest pion is absorbed by
 the off-shell nucleon what corresponds to intra-deuteron distances of the
 order of $\sim 1/\sqrt{m\mu}\simeq 0.6 fm$. This means that the absorption
 process should be sensitive to the dynamics of the $\pi NN$ system at small
 distances. In this paper we concentrate mainly on the investigation of the
 role of these effects and the contribution of the $P$-wave of the deuteron
 wave function \ci{gross,gross2}. The sensitivity of all the polarization
 observables to these effects is studied, and it is shown that some
 polarization characteristics can change the sign by including the off-mass
 shell effects of nucleons inside the deuteron.
 The detailed covariant formalism of the construction of the relativistic
 invariant amplitude of the reaction $NN\to d\pi$ and the helicity amplitudes
 for this process are presented in chapter 2.
 We analyse in detail both the pole graph, one-nucleon exchange, and the
 triangle diagram, i.e. the pion rescattering graph, in sections 3.
 The inputs by this consideration, the covariant pseudoscalar $\pi NN$  and
 deuteron $d\to pn$ vertices, are discussed in detail. The discussions of
 the obtained results and the comparison with the experimental data are
 presented in chapter 5. At least the conclusion is presented in the last
 section 6.

 \section{\bf General Formalism}

 $\bullet~~${\bf \it Relativistic invariant expansion of the amplitude }

 We start with the basic relativistic expansion of the reaction amplitude
 $NN\to d\pi$ using Itzykson-Zuber conventions \ci{zuber}.
 In the general case, the relativistic amplitude of the production of two
 particles of spins $1$ and $0$ by the interaction of two spin $1/2$
 particles has $6$ relativistic invariant amplitudes if all particles are
 on-mass shell and taking $P$-invariance into account.
 It can be written in the following form:
\begin{equation}
\hspace{-1cm}
\begin{minipage}{5.2cm}
\begin{center}
\unitlength1cm
\begin{picture}(6,4)
\thicklines
\put(3,2){\circle{1.4}}
\put(0,0){\makebox(6,4){${\cal X}_{\mu}$}}
\thinlines
\multiput(1,1.5)(0,1){2}{\vector(1,0){0.8}}
\multiput(1.8,1.5)(0,1){2}{\line(1,0){0.7}}
\multiput(3.5,1.5)(0.3,0){4}{\line(1,0){0.15}}
\put(4.7,1.5){\vector(1,0){0.15}}
\thicklines
\put(3.5,2.51){\line(1,0){1.265}}
\put(3.55,2.45){\line(1,0){1.21}}
\put(4.6,2.385){\makebox(0.2,0.1)[lb]{$>$}}
\put(0.5,2.8){\parbox[b]{2cm}{$\bar v_{\sigma_2}(p_2)$}}
\put(0.5,1){\parbox[b]{2cm}{$u_{\sigma_1}(p_1)$}}
\put(4.5,2.8){\parbox[b]{2cm}{$e^{~(\beta)}_\mu(d)$}}
\put(4.8,1.05){\parbox[b]{2cm}{$\varphi_\pi$}}
\end{picture}
\end{center}
\end{minipage}
\hfill
 \CM_{\gs_2,\gs_1}^{\beta}(s,t,u)=\bar v_{\gs_2}(p_2)\chi^\mu(s,t,u)
                           u_{\gs_1}(p_1)e^{~(\beta)}_\mu(d)\gvp_\pi
\label{IA1}
\end{equation}
 where $u_{\gs_1}$ and  $\bar v_{\gs_2}$ are the spinor and anti-spinor
 of the initial nucleons with spin projections $\gs_1$ and $\gs_2$,
 respectively; $e_\mu(d)$ is the deuteron polarization vector,
 $\gvp_\pi$ is the $\pi$-meson field; $s,t,u$ are the invariant
 variables determined in Appendix I.

 For example, for the $pp\to d\pi^+$ process, this amplitude should be
 symmetrized over the initial proton states, and therefore it takes the form:
 \be
 \bar \CM_{\gs_2,\gs_1}^{\beta}=\frac{1}{\sqrt{2}}\left[
 \CM_{\gs_2,\gs_1}^{\beta}(s,t,u)+
        {(-1)}^\beta\CM_{\gs_1,\gs_2}^{\beta}(s,u,t)\right]
 \label{IA2}
 \ee
 The second term in (\ref{IA2}), corresponding to the exchange of two protons,
 is equivalent to the exchange of the $t-$ and $u-$ variables.

 The amplitude $\chi_\mu$ for the process  $NN\to d\pi$
 can be expanded over six independent amplitudes \ci{zuber}:
 \be
 \chi_\mu=\g_5\left(\CX_1\g_\mu+\CX_2{p_\mu\over m}+\CX_3{p'_\mu\over m}
         +\CX_4{p_\mu\over m}\widehat{p'\over m}
         +\CX_5\left(\g_\mu\widehat{p'\over m}-\widehat{p'\over m}\g_\mu\right)
         +\CX_6{p'_\mu\over m}\widehat{p'\over m}\right).
 \label{IA3}
 \ee

 $\bullet~~${\bf \it Helicity formalism.}

 To calculate the observables, differential cross sections and
 polarization characteristics, it would be very helpful to construct the
 helicity amplitudes of the considered process $NN\to d\pi$. So, we use for 
 this reaction the helicity formalism presented in Ref.\ci{soffer}.
 Let us introduce initial nucleon helicities $\mu_1,\mu_2$
 and the final deuteron $\lambda$, and helicity amplitudes
 $\bar\CM^\lambda_{\mu_2,\mu_1}(W,\vartheta)$ depending on initial
 energy $W$ in the $N-N$ c.m.s. and scattering angle $\vartheta$
 analogous to \ci{locher}. This amplitude
 $\bar\CM^\lambda_{\mu_2,\mu_1}(W,\vartheta)$ corresponds to the transition
 of the $NN$ system from the state with helicities  $\mu_1,\mu_2=\pm 1/2$
 to the state with $\lambda=\pm 1,0$.

 With respect to discrete symmetries, we have from parity conservation 
 \ci{soffer}:
 \be  \CM_{\mu_2\mu_1}^\lambda=
 \eta_P(-1)^{(\mu_2-\mu_1)-\lambda}\CM_{-\mu_2-\mu_1}^{-\lambda}=
 (-1)^{\mu_2+\mu_1+\lambda}\CM_{-\mu_2-\mu_1}^{-\lambda}.
 \label{HA4}
 \ee
 Time - reversal symmetry leads to
 \be
 \CM_{\mu_2\mu_1}^\lambda=
 (-1)^{(\mu_2-\mu_1)-\lambda} \CM^{\mu_2\mu_1}_{\lambda},
 \label{HA5}
 \ee
 where
 $\eta_P={\eta_1\eta_2\over\eta_\pi\eta_d}(-1)^{s_d-s_1-s_2}=(-1)$;
 $\eta_i,s_i$ are internal parities and spins of particles.

 Introducing $6$ helicity amplitudes as \ci{locher}
 \be
 \Phi_{^1_3}=\bar\CM_{++}^\pm;~\Phi_{^2_5}=\bar\CM_{+\pm}^0;~
 \Phi_{^4_6}=\bar\CM_{+-}^\pm;
 \label{HA2}
 \ee
 which satisfy the following symmetry equations
 \be
 \Phi_{1,3}(\vartheta)&=&-\Phi_{1,3}(\pi-\vartheta);  \nonumber\\
 \Phi_{2,5}(\vartheta)&=&\Phi_{2,5}(\pi-\vartheta);  \nonumber\\
 \Phi_{4,6}(\vartheta)&=&\Phi_{6,4}(\pi-\vartheta)~,
 \label{HA3}
 \ee
 one can calculate all the observables over a range of $0<\gvt<\pi/2$.

 All the amplitudes $\bar\CM_{\mu_2\mu_1}^\lambda(W,\vartheta)$ should 
 vanish at forward and backward angles, and therefore we use the
 amplitudes introduced by Ref.\ci{soffer}:
 \be
 \bar\CM_{\mu_2\mu_1}^\lambda(W,\vartheta)=
 \left(Sin{\gvt\over2}\right)^{|\mu+\gl|}
 \left(Cos{\gvt\over2}\right)^{|\mu-\gl|}
 \widetilde\CM_{\mu_2\mu_1}^\lambda(W,\vartheta),
 \label{HA7}
 \ee
 where $(\mu=\mu_1-\mu_2)$ and
 $\widetilde\CM_{\mu_2\mu_1}^\lambda(W,\vartheta)$ are the non-vanishing
 amplitudes at $\gvt=0$ and $\gvt=\pi$.

Let us present now the relation of helicity amplitudes $\{\Phi_i\}_{i=1}^{6}$
to the invariant functions $\{\CX_i\}_{i=1}^{6}$.
We choose the axis $z$ along the nucleon momentum $\vec p_1$.
Using expansion (\ref{IA3}), one can get the following form of the
helicity amplitudes:
 \be
 \widetilde\Phi_{^1_3}&=&\pm\sqrt2\left\{{p\over m}\left({\gve\over m}\CX_2^a+
 {\gve_d-\gve\over m}\CX_4^a\right)-2{p(\gve_d-\gve)\mp k\gve\over m^2}\CX_5^a
 \right\}~,
 \nonumber\\
 \widetilde\Phi_2&=&-{k\over M}\left\{\CX_1^s-{\gve\over m}
 \left({\gve\over m}\CX_3^s+{\gve_d-\gve\over m}\CX_6^s\right)\right\}
 \nonumber\\
 &-&{p\over m}\left\{{\gve_d\over M}\left({\gve\over m}\CX_2^a+
 {\gve_d-\gve\over m}\CX_4^a\right)
 +2{\gve\gve_d-M^2\over Mm}\CX_5^a\right\}Cos\gvt~,
 \nonumber\\
 \widetilde\Phi_{^4_6}&=&\sqrt2\left\{\left({p\over m}\CX_1^s\mp2{k\over m}\CX_5^a\right)\mp
 2{p^2k\over m^3}
 \left(\begin{array}{c}Sin^2\gvt/2\\Cos^2\gvt/2\end{array}\right)\CX_4^a\right\}~,
 \nonumber\\
 \widetilde\Phi_5&=&2{p\over m}\left\{{\gve_d\over M}\left(\CX_1^s
 +{pk\over m^2}\CX_4^aCos\gvt\right)-{\gve k^2\over Mm^2}\CX_6^s\right\}~.
 \label{HA8}
 \ee
 where $\CX_i^{\{_a^s\}}(s,t,u)$ are symmetric and antisymmetric combinations
 $\CX^{\{_a^s\} }=(\CX_i(\gvt)\pm\CX_i(\pi-\gvt))/\sqrt2$.
 All symmetry properties (\ref{HA3}) are satisfied by these amplitudes. 
 
 The helicity amplitudes are decomposed into partial waves by
 (see \ci{locher})
 \be
 \bar\CM_{\mu_2\mu_1}^\lambda(W,\gvt)=
 \sum_J{2J+1\over2}\left(\CM ^J(W)\right)_{\mu_2\mu_1}^\gl d_{\mu,-\gl}^J(x)~;
 \label{HA10}
 \ee
 where $x=Cos\gvt$, and the  azimuthal angle $\gvp$ is taken to be zero.
 Using orthogonality relations for the $d-$function, one obtains
 \be
 \left(\CM^J(W)\right)_{\mu_ 2\mu_1}^\gl=\int_{-1}^1d_{\mu,-\gl}^J(x)
 \bar\CM_{\mu_2\mu_1}^\lambda(W,\gvt)dx
 \label{HA11}
 \ee
 and from symmetry relation (\ref{HA2}) one can find that
 $\Phi^J_{^4_6}=(-1)^{J+1}\Phi^J_{^6_4}$.

 \section{Reaction Mechanism}

$\bullet~~$ {\bf \it  One-nucleon exchange (ONE) and $\pi NN$-vertex.}

 Within the framework of the one-nucleon exchange model, the amplitude
 $\chi_\mu$ can be written in a simple form:
\begin{equation}\begin{array}{cc}
\begin{minipage}{5cm}
\begin{center}
\unitlength1cm
\begin{picture}(5,3)
\multiput(1,1)(0,1.2){2}{\vector(1,0){1.1}}
\multiput(2.1,1)(0,1.2){2}{\line(1,0){0.9}}
\multiput(3,1)(0.3,0){6}{\line(1,0){0.15}}
\put(4.8,1){\vector(1,0){0.2}}
\put(3,1){\circle*{0.1}}
\put(3,2.23){\line(1,0){2}}
\put(3,2.17){\line(1,0){2}}
\put(4.8,2.1){\makebox(0.2,0.1)[lb]{$>$}}
\put(3,2.2){\circle*{0.25}}
\thinlines
\put(3,1){\vector(0,1){0.7}}
\put(3,1.7){\line(0,1){0.5}}
\put(1,2.4){\parbox[b]{1cm}{$p_2$}}
\put(1,0.7){\parbox[b]{1cm}{$p_1$}}
\put(4.7,2.4){\parbox[b]{1cm}{$d$}}
\put(4.7,0.7){\parbox[b]{1cm}{$\pi$}}
\put(3.1,1.5){\parbox{1cm}{$n$}}
\put(2.8,2.5){\parbox[b]{1cm}{$\bar\Gamma_\mu$}}
\put(2.8,0.6){\parbox[b]{1cm}{$\Gamma_\pi$}}
\end{picture}
\end{center}
\end{minipage}
&
 \chi_\mu=g^+h_N(n^2)\bar\gG_\mu(d)\CS_\CF(n)\gG_\pi~,
 \label{ONE1}
\end{array}\end{equation}
 where
 $\bar\gG_\mu(d)$ is the deuteron vertex $pn\to d$ with one off-mass
 shell nucleon, $\CS_\CF(n)=\left(\widehat n-m+i0\right)^{-1}$ is the
 fermion propagator; the value of the coupling constant is
 $g^+=\sqrt2g^0=\sqrt2g~,~~g^2/4\pi=14.7~$, and $n^2=(d-p_2)^2=t$;
 the function $h_N(n^2)$ describes the vertex $NN\pi$ where one nucleon
 is an off-mass shell, but the other one and the pion are on-mass shell
 particles \ci{locher}. The vertex $\bar\gG_\mu(d)$ can be related to
 the deuteron wave function ($\CD\CW\CF$) with the help of the following
 equation \ci{gun1,gun2}:
 \be
 \bar\Psi_\mu={\bar\gG_\mu\over n^2-m^2+i0}
             =\gvp_1(t)\g_\mu+\gvp_2(t){n_\mu\over m}+
 \left(\gvp_3(t)\g_\mu+\gvp_4(t){n_\mu\over m}\right){\widehat n-m\over m}~.
 \label{ONE2}
 \ee
 The formfactors $\gvp_i(t)$ are related to two large components of
 the $\CD\CW\CF$ $u$ and $w$ (corresponding to the $^3\CS_1$ and $^3\CD_1$
 states) and to small components $v_t$ and $v_s$ (corresponding to the
 $^3\CP_1$ and $^1\CP_1$ states) as in \ci{gross}.

 Let us discuss now the problem connected with the form of the $\pi NN$
 vertex $\gG_\pi$. In the general case, it can be expanded over four
 covariant quantities if all particles are an off-mass shell \ci{kazes}:
 \be
 \gG_\pi=\gamma_5(F_1 + F_2 \frac{\widehat p_f + m}{m} +
 F_3\frac{\widehat p_i - m}{m} + F_4\frac{\widehat p_f + m}{m}
 \frac{\widehat p_i - m}{m})~;
 \label{ONE4}
 \ee
 here $p_i,p_f$ are the four-momenta of initial and final nucleons,
 $\{F_i(t; p_i^2, p_f^2)\}_{i=1}^4$ are some functions depending on the
 relativistic invariant transfer $t=(p_i-p_f)^2$ and their masses
 $p_{i,f}^2$ or the so-called pion formfactors. In our case, one nucleon
 $(p_f\equiv n)$ is the off-mass shell only, and therefore we have two terms 
 in eq.(\ref{ONE4}) instead of four because the third and the fourth ones are
 vanishing, taking into account the Dirac equation for a free fermion.
 Then, eq.(\ref{ONE4}) can be written in the form:
 \be
 \gG_\pi=\gamma_5\left(F_1(t)+F_2(t)\frac{\widehat n+m}{m}\right)~=
 \gl F^{\CP\CS}(t)\g_5+(1-\gl)F^{\CP\CV}(t)\g_5{\widehat\pi\over2m}~,
 \label{ONE5}
 \ee
 
 Note, according to the so-called equivalence theorem \ci{shweb} the
 sum of all Born graphs for elementary processes, for example the 
 pion photoproduction on a nucleon and the other ones, is invariant
 under chiral transformation. This means that starting with the 
 Lagrangian appropriate to the pseudoscalar $(\CP\CV)$ coupling, one
 ends up in the Lagrangian appropriate to the pseudoscalar 
 $(\CP\CS)$ coupling by performing a chiral transformation. This 
 equivalence theorem is related to the processes for elementary
 particles. But in our case, for the reaction $NN\rightarrow d\pi$
 there is a bound state, a deuteron, and therefore reducing this process
 to the one where only elementary particles participate, we will have
 the diagrams of a higher order over the coupling constant
 than the Born graph. So, the equivalence theorem cannot be applied 
 to our considered processes. Therefore, the vertex $\Gamma_\pi$
 in our case can be written in the form of eq.(\ref{ONE5})
 which is actually a linear combination of pseudoscalar and pseudovector
 coupling with the so-called mixing parameter $\gl$.

 For the on-mass shell neutron $(n^2=m^2)$ and the virtual pion, we have
 $\gG_\pi=\g_5$. Finally, using equations (\ref{IA3}), (\ref{ONE2}) and
 (\ref{ONE5}) ($n=p'+p$), one can find the following forms of the invariant
 amplitudes $\{\CX_i\}_{i=1}^6$
 \be
 \CX_1&=&g^+\left\{F_2\gvp_1-2(F_1+F_2)\gvp_3\right\}{t-m^2\over2m}~;
 \nonumber\\
 \CX_2&=&g^+m\left[F_1(\gvp_1+\gvp_2)-\left\{F_2(\gvp_2+\gvp_3-\gvp_4)
 -2F_1\gvp_4\right\}{t-m^2\over2m^2}\right]~;         \nonumber\\
 \CX_3&=&g^+m\left[F_1(2\gvp_1+\gvp_2)-\left\{F_2
 (\gvp_2+2\gvp_3-\gvp_4)-2F_1\gvp_4\right\}{t-m^2\over2m^2}\right]~;
 \nonumber\\
 \CX_4&=&\CX_6~=~-g^+m\left\{F_1\gvp_2-F_2{t-m^2\over2m^2}\gvp_4\right\}~;
 \nonumber\\
 \CX_5&=&g^+{m\over2}\left\{F_1\gvp_1-F_2{t-m^2\over2m^2}\gvp_3\right\}~.
 \label{ONE9}
 \ee
 Note, the amplitudes $\{\CX_i\}_{i=1}^6$ satisfy the following equations:
 $\CX_2-\CX_3+2\CX_5=0~;~~\CX_4=\CX_6~$.

 The $dNN$ vertex has been studied by Buck and Gross \ci{gross} within the
 framework of the Gross equation of nucleon-nucleon scattering. They used
 a one boson exchange (OBE) model with $\pi, \rho, \go$ and $\gs$ exchange.
 In their study, they suggest that the formfactors $F^{\CP\CS}$ and
 $F^{\CP\CV}$ have the same $t$ - dependence, in particular
 $F^{\CP\CS}(t)=F^{\CP\CV}(t)=h_N(t)$, and consider
 $\gl=0.0; 0.2; 0.4; 0.6; 0.8$ and $1.0$. In each case, the parameters of the
 OBE model were adjusted to reproduce the static properties of the deuteron.
 They found that the total probability of the small components of the
 $\CD\CW\CF$: $P_{small}=\int_0^\infty p^2dp\left[v_t^2(p)+v_s^2(p)\right]$,
 increases monotonically with growing $\gl$ from approximately $0.03\%$ for
 $\gl=0$ to approximately $1.5\%$ for $\gl=1$.

 The function $h_N(t)$ is the nucleon formfactor caused by the virtual
 nucleon and can be taken by the Breit-Wigner-type form suggested by
 \ci{nut} and \ci{locher}.\\
 \be
 h_N(t)={m-E_R\over \sqrt t-E_R+i\gG(t)/2}~,~~
 \gG(t)=2\bar\ga\gT(\sqrt t-m-\mu)
 exp\left\{-{\bar\gb\over\sqrt t-m-\mu}\right\}
 \label{ONE7}
 \ee
 at $\bar\ga=0.26~GeV~;~~\bar\gb=0.40~GeV~;~~E_R=1.42~GeV$.

 Let us analyse the contributions of functions $\gvp_1,\gvp_2$ determining
 the form of $\CD\CW\CF$ to the invariant amplitudes $\CX_i$.
 It is interesting to consider the case when the off-mass shellness of the
 nucleon is small, e.g. $\gvp_3=\gvp_4=0$ and $\gl=1$. In this case,
 we have $5$ relativistic invariant amplitudes $\{\CX_i\}_{i=2}^6$ instead
 of $6$.  \\

$\bullet~~$ {\bf \it  Second-order graphs}

Let us consider now the second order graph corresponding to the
rescattering of the virtual $\pi$-meson by the initial nucleon.
This mechanism of the $NN\to \pi d$ process has been analysed by many
authors, see, for example, \ci{barry,locher}. Our procedure of the
construction of the helicity amplitudes corresponding to the
triangle graph is different from the ones published by \ci{barry,locher},
and so we present the proof of the forms of these amplitudes briefly.
\be
\begin{minipage}{6cm}
\begin{center}
\unitlength1cm
\begin{picture}(5,3)
\put(0,2.2){\vector(1,0){0.8}}
\put(0.8,2.2){\line(1,0){0.7}}
\put(1.5,2.2){\vector(1,0){1.3}}
\put(2.8,2.2){\line(1,0){1.1}}
\put(1.5,2.2){\line(1,-1){0.36}}
\put(1.9,1.8){\vector(1,-1){0.36}}
\put(2.3,1.4){\line(1,-1){0.36}}
\put(2.7,1){\vector(1,1){0.7}}
\put(3.2,1.5){\line(1,1){0.7}}
\put(1.5,2.2){\circle*{0.1}}
\thicklines
\put(3.9,2.23){\line(1,0){1.5}}
\put(3.9,2.17){\line(1,0){1.5}}
\put(5.2,2.1){\makebox(0.2,0.1)[lb]{$>$}}
\put(3.9,2.2){\circle*{0.25}}
\thinlines
\put(0,1){\vector(1,0){0.8}}
\put(0.8,1){\line(1,0){1.9}}
\multiput(2.7,1)(0.45,0){5}{\line(1,0){0.35}}
\put(5,1){\vector(1,0){0.37}}
\put(2.7,1){\circle*{0.15}}
\put(0.1,2.4){\parbox[b]{1cm}{$p_2$}}
\put(0.1,0.7){\parbox[b]{1cm}{$p_1$}}
\put(5,2.4){\parbox[b]{1cm}{$d$}}
\put(5,0.7){\parbox[b]{1cm}{$\pi$}}
\put(2.6,2.4){\parbox{1cm}{$\eta$}}
\put(3.5,1.5){\parbox{1cm}{$k$}}
\put(1.7,1.5){\parbox{1cm}{$q$}}
\put(1.3,2.45){\parbox[b]{1cm}{$\Gamma_\pi$}}
\put(3.7,2.45){\parbox[b]{1cm}{$\bar\Gamma_\mu$}}
\put(2.4,0.5){\parbox[b]{1cm}{$f_{\pi N}^{el}$}}
\end{picture}
\end{center}
\end{minipage}
\hfill
\chi_\mu^{sp}=
{g^+\over (2\pi)^3}\int h_\pi(q^2){\CF_\mu\left(\vec\eta,
\eta_0=\sqrt{\vec\eta^2+m^2}\right)\over q^2-\mu^2}{d^3\eta\over2\eta_0}
\label{SO1}
\ee
where $h_\pi(q^2)$ is the pion formfactor corresponding to the off-mass
shell $\pi$-meson in the intermediate state; its form has been chosen
in the monopoly one $h_\pi(q^2)=(\gL^2-\mu^2)/(\gL^2-q^2)$
as like as in \ci{barry,2}; here $\gL$ is the corresponding cut-off
parameter. The general form of $\CF_\mu$ can be written as follows:
\be
&&\CF_\mu=\gG_\pi\CS_\CF^c(\eta){\bar\gG}_\mu\CS_\CF(k) f^{el}_{\pi N}~,
\label{SO2}
\ee
where $f^{el}_{\pi N}$ is the amplitude of $\pi N$ elastic scattering;
it can be presented as expansion over two off-shell invariant amplitudes
$f^{el}_{\pi N}=(A+B\widehat\pi)$ which depend on four momenta. We compute
A and B from the on-shell $\pi N$ partial wave amplitudes
$\CT_{l\pm}^{on}(s_{\pi N})$ under the assumption
\be
\CT_{l\pm}(s_{\pi N},t_{\pi N}, u_{\pi N})\approx\CT_{l\pm}^{on}(s_{\pi N})~,
\label{SO3}
\ee
where $\CT_{l\pm}^{on}(s_{\pi N})$ are taken from the Karlsruhe-Helsinki
phase shift analysis \ci{holer}. However, in the partial wave decomposition
of the invariant functions, full off-shell angular momentum projectors
are used for the lowest waves in the manner discussed for the $NN\to NN\pi$
reaction in Ref.\ci{kroll}.

Using the covariant form of the deuteron wave function $\bar\Psi_\mu$
(\ref{ONE2}), the matrix $\CF_\mu$ (\ref{SO2}) can be decomposed into a
suitable set of invariant functions :
\be
\CF_\mu=m^2\sum_i\widehat\CI_\mu(i)a_i~;
\label{SO5}
\ee
the matrices $\widehat\CI_\mu(i)$ and functions $a_i$ are presented
in Appendix III.

We are faced with a three dimensional integro - operator over the loop
momentum.
\be
\widehat\CI^{sp}={g^+\over(2\pi)^2}\int_0^{\eta_m}
{\eta^2d\eta\over2\sqrt{\eta^2+m^2}}\int_{-1}^1
{h_\pi(q^2)dCos\gvt_\eta\over q^2-\mu^2}\int_0^{2\pi}d\gvp_\eta~.
\label{SO6}
\ee
The square of energy $s_{\pi N}$, the momentum transfer $u_{\pi N}$ and the 
square of virtual pion mass $q^2$ do not depend on azimuth $\gvp_\eta$:
\be
&&s_{\pi N}=(p_1+q)^2=s-2\sqrt s\eta_0+m^2~;
\nonumber \\
&&u_{\pi N}=(p_1-q)^2=t~;
\nonumber \\
&&q^2=2\left(m^2-\gve\eta_0-p\eta_3\right)~.
\label{SO7}
\ee
Note, at $T_p=0.578~GeV$ we have:
\be
\eta_m^0={s-(m+\mu)^2+m^2\over2\sqrt{s}};~
\eta_m=\sqrt{(\eta_m^0)^2-m^2}\approx0.366~GeV~.
\label{SO8}
\ee
In this kinematic region, the square of the pion four-vector $q^2$ is
space-like and the pion is moderately far from its mass shell 
$[0>q^2>-0.3~GeV^2]$ whereas an active nucleon is close to its mass shell
$[0.83>k^2>0.63~GeV^2].$

Triple integral (\ref{SO6}) over azimuth $\gvp_\eta$, polar angle
$\gvt_\eta$ and the magnitude of three-momentum $\eta$ must be done
numerically for which we used a Gauss method. There are 6 triple integrals
over a complicated complex integrand for each scattering angle.

\section{\bf Observables }

Using the helicity amplitudes discussed in section 2, one can calculate
all the observables: differential cross section, asymmetry, deuteron
tensor polarization and so on.

It is convenient to introduce hybrid reaction parameters $(\CH\CR\CP)$ for
the reaction $NN\to d\pi$ as \ci{locher,soffer,foroughi}
\be
(\ga\gb\mid LM)^{Mad}=\gve_\gb(-1)^M Tr\left[
\gs_\ga\gs_\gb\stackrel{+}{\CM}^\gl_{\mu_2\mu_1}T_M^L(s_d)\CM^\gl_{\mu_2\mu_1}
\right]\gS^{-1}~.
\label{Ob}
\ee
with $\gve_0=\gve_x=1;~\gve_y=\gve_z=1$, $\gs_\ga$ and $\gs_\gb$
$(\ga,\gb=0,x,y,z)$ the Pauli spin operators for initial nucleons and
$T_M^L(s_d)$ the spin-one tensor of rank $L\leq2$. The normalization of
the $\CH\CR\CP$ is such that $(00\mid00)=1$. Then, the differential cross
section is related to $\gS$ as
\be
\gS=2\sum_1^6{\mid \Phi_i\mid}^2=4\frac{p}{k}
{(\frac{m}{4\pi\sqrt{s}})}^{-2}\frac{d\sigma}{d\Omega}=
{1\over\gs_0}\frac{d\sigma}{d\Omega}~,
\label{Ob0}
\ee
where $p$ and $k$ are the momenta of initial proton and final deuteron
in the c.m.s.
There are $4\times4\times9=144$ $\CH\CR\CP$. However, since parity invariance
reduces the number of independent amplitudes to six, there are only $36$
linearly independent bilinear observables. They have the following symmetry
properties and relations:
\be
&&(\ga\gb\mid LM)~~is~~\left\{\begin{array}{c}real\\imag.\end{array}\right\}~~
  ~~if~~n_0+n_y+L~~is~~\left\{\begin{array}{c}even\\odd\end{array}\right\}~,
\nonumber \\
&&n_{0,y}~~is~~number~~of~~\gs_{0,y}~;~~
\nonumber \\
&&(\ga\gb\mid LM)_{\gvt}=(-1)^M(\gb\ga\mid LM)_{\pi-\gvt}~;
\nonumber \\
&&(\ga\gb\mid LM)=\gz_\ga\gz_\gb(-1)^{L+M}(\ga\gb\mid L-M)~,
\nonumber \\
&&\gz_0=\gz_y=1~;~~\gz_x=\gz_z=-1~.
\label{Ob1}
\ee
Let us present now the expressions for the following observables in the c.m.s.
using $\Phi_i$:
\begin{eqnarray}
A_{y0}& = &4Im(\Phi_1\Phi^*_4+\Phi_2\Phi^*_5+\Phi_3\Phi^*_6)\gS^{-1},~~~
A_{0y}(\gt)=A_{y0}(\pi-\gt),
\nonumber\\
A_{xz}& = &-4Re(\Phi_1\Phi^*_4+\Phi_2\Phi^*_5+\Phi_3\Phi^*_6)\gS^{-1},~~~
A_{zx}(\gt)=A_{xz}(\pi-\gt),
\nonumber \\
A_{zz}& = &-1+4(|\Phi_4|^2+|\Phi_5|^2+|\Phi_6|^2)\gS^{-1},
\nonumber \\
A_{yy}& = &-1+2(|\Phi_1+\Phi_3|^2+|\Phi_4+\Phi_6|^2)\gS^{-1},
\nonumber \\
A_{xx}& = &A_{zz}+2(|\Phi_1+\Phi_3|^2-|\Phi_4+\Phi_6|^2)\gS^{-1}.
\label{Ob2}
\end{eqnarray}
The expressions for the deuteron tensor polarization components are
the following:
\begin{eqnarray}
iT_{11} & = &-\sqrt{6}Im\left [(\Phi^*_1-\Phi^*_3)\Phi_2+
(\Phi^*_4-\Phi^*_6)\Phi_5 \right]\gS^{-1},
\nonumber \\
T_{20} & = &\left[1-6(|\Phi_2|^2+|\Phi_5|^2)\gS^{-1}\right]/\sqrt{2},
\nonumber \\
T_{21} & = &\sqrt{6}Re\left[(\Phi^*_1-\Phi^*_3)\Phi_2+
(\Phi^*_4-\Phi^*_6)\Phi_5\right]\gS^{-1},
\nonumber \\
T_{22} & = &2\sqrt{3}Re(\Phi^*_1\Phi_3+\Phi^*_4\Phi_6)\gS^{-1}=
(1+3A_{yy}-\sqrt{2}T_{20})/(2\sqrt{3})~.
\label{Ob3}
\end{eqnarray}

\section{Results and Discussions}

In order to investigate the effect of small components of the $\CD\CW\CF$,
we have calculated the differential cross section $d\sigma/d\Omega$,
polarization characteristics $A_{ii}, A_{y0}$, etc. for $pp\to d\pi^+$
as a function of scattering angle at proton kinetic energy $T_p=578 MeV$
corresponding to pion kinetic one $T_\pi=147 MeV$ because at this 
energy the probability of $\gD$-isobar production by the two - step 
mechanism is rather sizeable. All the calculated quantities are in the 
Madison convention and compared with the experimental data \ci{loch1,arndt} 
and partial-wave analysis ($\CP\CW\CA$) by R. A. Arndt et al. \ci{said} 
(dotted curve). The cut-off parameter $\gL$ and the mixing one $\gl$ 
corresponding to the $\pi NN$ vertex are chosen by the best fitting of the 
experimental cross section $d\gs/d\gO$ data. We have checked that the 
polarization curves change very little if we vary the cut-off parameter $\gL$.

Note that the contribution of the triangle graph is very large at
intermediate initial kinetic energies and much smaller at lower energies.
It is caused by a large value of the cross section of elastic $\pi N$
scattering because of a possible creation of the $\gD$-isobar at this energy.
One can stress that the application of Locher's form $\CD\CW\CF$ \ci{loch2}
does not allow one to reproduce the absolute value of the differential cross
section (see Fig.(1)) over the whole region of scattering angle $\gvt$.
But using the Gross approach for the $\CD\CW\CF$, one can describe
$d\gs/d\gO$ at $\gl=0.6-0.8$ rather well.

The next interesting result which can be seen from Fig.(1) is a large
sensitivity of all the polarization characteristics to the small components 
of the $\CD\CW\CF$. The asymmetry $A_{y0}$ and the vector polarization
$iT_{11}$ calculated within the framework of Gross's approach particularly
show this large sensitivity. These quantities are interference dominated and
sensitive to the phases. The results for $iT_{11}$ have a wrong sign
with Locher's form $\CD\CW\CF$ \ci{loch1}. On closer inspection, we observe
that the first term in eq.(\ref{Ob3}), $(\Phi^*_1-\Phi^*_3)\Phi_2$, is very
big due to constructive interference $\Phi_1\approx -\Phi_3$. It is caused by
the $N\triangle$ configuration in a relative $\CS$ wave having $pp$ spin zero
($^1\CD_2$ state). The $^1\CD_2$ partial-wave dominates making $\Phi_{1,2,3}$
large, but the results are the same contribution to $\Phi^{\CJ=2}_1$ and
$\Phi^{\CJ=2}_3$ (with opposite signs caused by the relevant Wigner
d-function signature). Since the contribution of $\Phi_{4,5,6}$ is
negligible, the sign problem for $iT_{11}$ is therefore very sensitive
to the $\Phi^{\CJ=0}_2$ (or $^1\CS_0$) partial wave. As $iT_{11}$ is very
nearly proportional to $\Phi_2$, the phase of $\Phi_2$ determines the sign
of $iT_{11}$.

The right structure of the observables starts to appear gradually in the
theoretical curves as one increases the mixing parameter $\gl$ in the
Buck-Gross model, that is to say, as one increases the probability of the
small components in the $\CD\CW\CF$. We have checked that this structure
originates indeed from the small components $v_t$ and $v_s$ in
eq.(\ref{ONE2}). If we make $v_t=v_s=0$ in the Buck-Gross model, then
all curves become very similar to Locher's ones. Similarly, if we vary
the $\pi NN$ vertex given by eq.(\ref{ONE5}) by considering $\gl$ between
$0$ and $1$ but keep Locher's $\CD\CW\CF$, then the curves change
very little again.

The proton spin correlations $A_{ii}$ are presented in Fig.(2). Actually, the
data on $A_{zz}$ is the measure of the $\Phi_{4,5,6}$ magnitudes because the
deviation of $A_{zz}$ from $-1$ is determined by these amplitudes (\ref{Ob2}).
According to the partial-wave decomposition, $\Phi_4$ and $\Phi_6$
are the amplitudes containing only triplet spin states in the $pp$ channel.
One can conclude that the magnitudes of the spin-triplet amplitudes are
somewhat small. As for $A_{yy}$ and $A_{xx}$, the terms proportional to
$\Phi_1+\Phi_3$ can be neglected because there is a phase relation
$\Phi_1\approx -\Phi_3$. Therefore, the deviation of $A_{yy}$ and $A_{xx}$
from $-1$ is determined by $\Phi_{4,6}$ again, whereas $\Phi_5$ does not
contribute to the numerator of $A_{yy}$.

One can also see a large sensitivity of the observables $A_{ii}$ to the used
form of $\CD\CW\CF$. The application of Gross's approach by the construction of
$\CD\CW\CF$ \ci{gross} results in the shapes of these characteristics which are
different from the corresponding ones obtained within the framework of Locher's
approach \ci{loch3}.

Note, the energy dependence of all the observables within the framework of the
suggested approach is the subject of our next investigation.

\section{Summary and Outlook}

A relativistic model for the reaction $NN\to d\pi$ has been discussed in
detail using two forms of the $\CD\CW\CF$ \ci{loch1} and \ci{gross}. One of
them \ci{loch1} was already used in the analysis of the $pp\to d\pi$ process
\ci{locher} also taking into account the two-step mechanism with a virtual
pion in the intermediate state. The difference between our approach and the
model considered in \ci{locher} is the following. We have analysed the
sensitivity of all the observables to the form of $\pi NN-$current and the
choice of the $\CD\CW\CF$ relativistic form. First of all, from the results
presented in Fig.(1,2), one can see very large sensitivity of all the
observables, especially of the polarization characteristics to the choice
of the $\CD\CW\CF$ form. The inclusion of the $P$-wave contribution in the
$\CD\CW\CF$ within the framework of Gross's approach \ci{gross} results in a
better description of the experimental data on the differential cross section
and the polarization observables. The next interesting result is related
to the extraction of some new information on the off-shell effects due
to a virtual (off-shell) nucleon. Comparing the observable with the
experimental data (see Fig.(1,2)), one can test the assumption, suggested
by \ci{gross2}, of a possible form of the pion formfactor and conclude that
one cannot use the mixing parameter $\gl=1$ as like as in \ci{loch1}.

One can stress that the one-nucleon exchange and the pion rescattering graphs
have been studied only in this paper in order to investigate
very important effects: off-mass shellness of nucleon and pion, and $P-$wave
contribution to the $\CD\CW\CF$. The interactions in the initial $NN$ and
final $d\pi$ states can be in principle contributed to the total amplitude 
of the considered reaction. However, it will be as a separate stage of this
study because a more careful inclusion of elastic $NN$ and $d\pi$
interactions at intermediate energies is needed.

Finally, let us stress that there is in principle an alternative approach
to study the $\CD\CW\CF$ at small distances based on the non - nucleon or
quark degree of freedom \ci{lyk,glozman,kobush}. However, the main goal of 
our paper is to show the role of the conventional nucleon degrees of freedom
in the deuteron by analising the processes of the type $NN\to d\pi$.
Therefore, we didn't analyse the application of quark approaches to this 
reaction.

 {\bf Acknowledgements.}

We gratefully acknowledge  very helpful discussions with
V. R. Pandharipande, R. Machleidt, S. Moszkovsky and E. A. Strokovsky.

 \section{\bf Appendix I}

$\bullet~~$ {\sl {\bf Kinematics of $NN\to d\pi$.}}

The $\CS$-matrix element of the reaction
 $\CS_{\gs_2\gs_1}^\beta =< \pi d, out | p_1 p_2, in >$
is related to the corresponding $\CM$-matrix element by the following
equation:
\be
 \CS_{\gs_2\gs_1}^\beta = {1\over (2\pi)^2}{m\over{\sqrt{\varepsilon_1
  \varepsilon_22\varepsilon_\pi2\varepsilon_d}}}
 \delta^4\left(\pi+d-p_1-p_2\right)\CM_{\gs_2\gs_1}^\beta,
\label{Ap2}
\ee
where $\beta, \gs_1$ and $\gs_2$ are the spin indices of deuteron
polarization and spin projections of initial nucleons.

As is well-known, Mandelstam's variables
\be
 s=(p_1+p_2)^2~~;~~t=(d-p_2)^2~~;~~u=(d-p_1)^2~~,
\label{Ap3}
\ee
are related by the condition: $ s+t+u=M^2+2m^2+\mu^2=h$.

Let us introduce the following variables:
\be
P=p_1+p_2~~~~~,~~~~~P^2=s~~~~~;\nonumber\\
p=(p_1-p_2)/2~~,~~~p'=(d-\pi)/2.
\label{Ap4}
\ee
Note,
\be
 p^2=m^2-{s\over4}\leq0~;~~p'^2={1\over2}(M^2+\mu^2)-{s\over4}\leq0~;~~
 p'^2+p^2={t+u\over2}~;  \nonumber\\
(pP)=0~;~~(p'P)={1\over2}(M^2-\mu^2)~;~~(p'p)\equiv\nu={t-u\over4}\leq0~~.
 ~~~~\label{Ap5}
\ee

Let us now introduce two space-time four-vectors orthogonal to $P$ and
$p$ :
\be
N_\mu={1\over\sqrt{-p^2P^2}}\gve_{\mu\nu\gr\gs}p'^\nu p^\gr P^\gs~;~~(Np')=0;
~~~~~~~~~~\nonumber\\
L_\mu={1\over\sqrt{-p^2P^2}}\gve_{\mu\nu\gr\gs}N^\nu p^\gr P^\gs=
\left(p'_\mu-{\nu\over p^2}p_\mu\right)-{(p'P)\over P^2}P_\mu~;
\nonumber\\
N^2=L^2=
\left\{\left(p'^2-{\nu^2\over p^2}\right)-{(p'P)^2\over P^2}\right\}\leq0~;
~~(LN)=0.~~~
\label{Ap6}
\ee
Then, one can get the whole system of orthogonal unit four-vectors
$\{e_\mu^{(\gs)}\}_{\gs=0}^3$,
three of them are space-like :
\be
e_\mu^{(1)}\equiv l_\mu={L_\mu\over\sqrt{-L^2}}~;~~
e_\mu^{(2)}\equiv n_\mu={N_\mu\over\sqrt{-N^2}}~;~~
e_\mu^{(3)}\equiv e_\mu={p_\mu\over\sqrt{-p^2}}~;~~
\label{Ap7}
\ee
and one of them is time-like:
\be
e_\mu^{(0)}={P_\mu\over\sqrt{P^2}}~.
\label{Ap8}
\ee
They satisfy the following conditions :
\be
g^{\mu\nu}e_\mu^{~(\gs)}e_\nu^{~(\gr)}=g^{\gs\gr}~;~~~~~
e_\mu^{~(\gs)}e_\nu^{~(\gr)}g_{\gs\gr}=g_{\mu\nu}~,~~~~~
\label{Ap9}
\ee
 Therefore, any four-vector $a_\mu$ can be expanded over this
unit orthogonal system, i.e.:
\be
a_\mu=\left(ae^{(\gs)}\right)e_\mu^{~(\gr)}g_{\gs\gr}~;~~
a^2=(ae)^2.
\label{Ap10}
\ee
For example, we can expand the matrix four-vector $\chi_\mu$ (\ref{IA1}) over
these basic vectors:
\be
\chi_\mu=\chi_ie_\mu^{(i)}=\chi_1l_\mu+\chi_2n_\mu+\chi_3e_\mu~,~~
\chi_i=-\chi^\mu e_\mu^{(i)}~=\g_5\left(a_i+b_i{\widehat l}\right);~
\label{IA4}
\ee

\section{\bf Appendix II}

$\bullet~~$ {\bf Pauli's representation of $NN\to d\pi$.}

In the c.m.s., we can use the next Pauli form of the reaction amplitude
\be
 \CM_{\gs_2\gs_1}^\beta(t,u) =
 w^+_{-\gs_2}\left(\vec\chi\vec e^{~(\beta)}\right) w_{\gs_1}.
\label{PR1}
\ee
The vector of the reaction is parametrized in the following form:
\be
\vec\chi=\sum_{i=1}^{6}\chi_i(\vec p^{~2},\vec n\vec p)\cdot \vec i=
   \chi_1\vec e_p+\chi_2\vec n+i\chi_3\left[\vec\gs\times\vec e_p\right]~
      \nonumber \\
    +i\chi_4\left[\vec\gs\times\vec n\right]
    +i\chi_5\left(\vec\gs\vec n\right)\left[\vec n\times\vec e_p\right]
    +i\chi_6\left(\vec\gs\vec e_p\right)\left[\vec n\times\vec e_p\right];
\label{PR2}
\ee
where:
$\vec e_p=\widehat{\vec p},~~\vec n=-\vec e_d~;~~
\vec e_p^{~2}=\vec n^{~2}=1~;~~z=(\vec n\vec e_p)~.$

And, finally, we have the following connection with invariant expansion
(\ref{IA3}):
\be
\chi_1&=&-{p\over m}\left({\varepsilon\over m}\CX_2+{\gve_d-\gve\over m}
       \left(\CX_4-2\CX_5\right)\right);
\nonumber\\
\chi_2&=&-{k\over M}\left(\CX_1-{\gve\over m}\left(
{\varepsilon\over m}\CX_3+{\gve_d-\gve\over m}\CX_6\right)\right)
\nonumber\\
&-&{p\over m}{\varepsilon_d-M\over M}\left({\varepsilon\over m}\CX_2
+{\gve_d-\gve\over m}(\CX_4-2\CX_5)+2{\varepsilon_d+M\over m}\CX_5\right)z;
\nonumber\\
\chi_3&=&{p\over m}\left({\varepsilon_d\over M}\left(\CX_1+{pk\over m^2}
z\CX_4\right)-{\varepsilon k^2\over Mm^2}\CX_6\right);
\nonumber\\
\chi_4&=&-{p\over m}\left({\varepsilon_d-M\over M}
\left(\CX_1+{pk\over m^2}z\CX_4\right)-{\gve k^2\over Mm^2}\CX_6\right)z
\nonumber\\
&-&{k\over m}\left({\gve^2\over m^2}\CX_4-\left(\CX_4-2\CX_5\right)\right);
\nonumber\\
\chi_5&=&-{p\over m}\left({\varepsilon_d-M\over M}
\left(\CX_1+{pk\over m^2}z\CX_4\right)-{\gve k^2\over Mm^2}\CX_6\right);
\nonumber\\
\chi_6&=&-{k\over m}{\varepsilon-m\over m}\left({\varepsilon\over m}\CX_4
+\left(\CX_4-2\CX_5 \right)\right).
\label{PR5}
\ee

 Helicity amplitudes (\ref{HA4}) can be related to the corresponding
 Pauli amplitudes $\{\chi_i\}_{i=1}^{6}$ (\ref{PR2}):
 \be
 \widetilde\Phi_{^1_3}&=&
 \sqrt2\left(\mp\chi_1^a-\chi_4^a+\chi_5^sCos\gvt+\chi_6^a\right)~;
 \nonumber\\
 \widetilde\Phi_2&=&\chi_1^aCos\gvt+\chi_2^s~;            \nonumber\\
 \widetilde\Phi_{^4_6}&=&\left\{\left(\chi_3^s\pm\chi_4^a\right)+2\chi_5^s
 \left(\begin{array}{c}Sin^2\gvt/2\\Cos^2\gvt/2\end{array}\right)\right\}~;
 \nonumber\\
 \widetilde\Phi_5&=&2\chi_3^sSin\gvt~.
 \label{HA9}
 \ee

\section{Appendix III}

In this appendix we give explicit expressions for helicity amplitudes
(\ref{HA4}) for the rescattering diagram. The evolution of the
expression for $\chi_\mu^{sp}$ (\ref{SO1}) is straightforward. The spin
structure operator $\CF_\mu$ (\ref{SO2}) here
\be
{1\over m^2}\CF_\mu=\sum_{i=1}^{16}a_i\widehat\CI_\mu(i)=
\gG_\pi{(-\widehat\eta+m)\over m}\bar\Psi_\mu{(\widehat k+m)\over m}
(A+B\widehat\pi)
\label{AIII1}
\ee
is a $4\times 4$ matrix in the spinor space and carries the label of deuteron
polarization. The first six of the operators $\widehat\CI_\mu(i)$ do
not depend on the integration variable:
\be
\widehat\CI_\mu(1)=\g_5\g_\mu ~;~~
\widehat\CI_\mu(2)=\g_5{p_\mu\over m} ~;~~
\widehat\CI_\mu(3)=\g_5{p'_\mu\over m} ~;~~
\nonumber \\
\widehat\CI_\mu(4)=\g_5\widehat\pi{p_\mu\over m} ~;~~
\widehat\CI_\mu(5)=\g_5\widehat\pi{p'_\mu\over m} ~;~~
\widehat\CI_\mu(6)=\g_5\g_\mu\widehat\pi ~.
\label{AIII2}
\ee
The next three of $\widehat\CI_\mu(i)$ depend only on $\eta :$
\be
\widehat\CI_\mu(7)=\g_5{\eta_\mu\over m}~;~~
\widehat\CI_\mu(8)=\g_5{\widehat\eta\eta_\mu\over m^2}~;~~
\widehat\CI_\mu(9)=\g_5\left(\g_\mu{\widehat\eta\over m}-
{\widehat\eta\over m}\g_\mu\right)~.
\label{AIII3}
\ee
The remaining $\widehat\CI_\mu(i)$ are:
\be
\widehat\CI_\mu(10)=\g_5{\widehat\eta p_\mu\over m^2} ~;~~
\widehat\CI_\mu(11)=\g_5{\widehat\eta p'_\mu\over m} ~;~~
\widehat\CI_\mu(12)=\g_5{\eta_\mu\over m}\widehat\pi~;~~
~~~~~\nonumber \\
\widehat\CI_\mu(13)=\g_5{\widehat\eta p_\mu\over m^2}\widehat\pi ~;~~
\widehat\CI_\mu(14)=\g_5{\widehat\eta p'_\mu\over m^2}\widehat\pi~;~~
\widehat\CI_\mu(15)=\g_5{\widehat\eta\eta_\mu\over m^2}\widehat\pi ~:~~
\nonumber \\
\widehat\CI_\mu(16)=\g_5\g_\mu{\widehat\eta\over m}\widehat\pi ~.
~~~~~~~~~~~~~~~~~~~~~~~~~~~~~~
\label{AIII4}
\ee

With little algebra, one finds
\be
a_1&=&A\left(2-{q^2\over m^2}\right)\gvp_1+B{m^2-t\over m}\gvp_1
+A{k^2-m^2\over m^2}\gvp_3~; \nonumber \\
a_2&=&-2A\gvp_1~=~a_3~;~~
a_4~=~-2B\gvp_1~=~a_5~;\nonumber \\
a_6&=&-{1\over m}A\gvp_1-B{q^2\over m^2}\gvp_1+B{k^2-m^2\over m^2}\gvp_3~;
\nonumber \\
a_7&=&-2A(2\gvp_1+\gvp_2)-A{q^2\over m^2}\gvp_2-B{m^2-t\over m}
\left(\gvp_1+\gvp_2\right)-A{k^2-m^2\over m^2}\left(\gvp_3+\gvp_4\right)~;
\nonumber \\
a_8&=&2A(\gvp_1+\gvp_2)+B{m^2-t\over m}\gvp_2+A{k^2-m^2\over m^2}\gvp_4~;
\nonumber \\
a_9&=&A\gvp_1+B{m^2-t\over2m}\gvp_1+A{k^2-m^2\over2m^2}\gvp_3~;
\nonumber \\
a_{10}&=&2A\gvp_1~=~a_{11}~;
\nonumber \\
a_{12}&=&{1\over m}A\left(2\gvp_1+\gvp_2\right)-2B\gvp_1-B{q^2\over m^2}\gvp_2
-B{k^2-m^2\over m^2}\left(2\gvp_3+\gvp_4\right)~;
\nonumber \\
a_{13}&=&2B\gvp_1~=~a_{14}~;
\nonumber \\
a_{15}&=&-{1\over m}A\gvp_2+2B\gvp_1+B{k^2-m^2\over m^2}\gvp_4~;
\nonumber \\
a_{16}&=&-{1\over m}A\gvp_1+B{k^2-m^2\over m^2}\gvp_3~.
\label{AIII5}
\ee

Calculating all the spinor matrix elements, one comes to the following
explicit expressions for the helicity amplitudes of the rescattering
diagram $(S=Sin\gvt;~C=Cos\gvt)$ :
{\small
\be
\chi_{^1_3}&=&\CM_{++}^\pm~=~\CM_{--}^\mp \nonumber \\
=\widehat\CI\left[\right.
&\pm&\left\{{p\over m}\left({\gve\over m}a_2+\gve_\pi a_4
+{\eta_0\over m}a_{10}\right)-{p\gve_\pi\pm k\gve\ov m}a_6\right\}{S\ov\sqrt2}
\nonumber \\
&-&{(e_\pm\eta)\ov m}\left\{{\gve\ov m}a_7+{\eta_0\over m}a_8+\gve_\pi a_{12}
\right\}-\sqrt2{\gve\left(\eta_1\pm i\eta_2C\right)\pm p\eta_0S\over m^2}a_9
\nonumber \\
&+&\left\{\eta_0{\gve\gve_\pi-pkC\over m^2}-{k\gve\ov m^2}
\left(\eta_1-i\eta_2\right)S+\eta_3{p\gve_\pi-k\gve C\ov m^2}\right\}
\left\{\pm{p\ov m}a_{13}{S\ov\sqrt2}-{(e_\pm\eta)\ov m}a_{15}\right\}
\nonumber \\
&-&\left\{{\gve_\pi\over m}{\left[(e_\pm\eta)+{\eta_1\pm i\eta_2C\ov\sqrt2}
\right]-\eta_0{k\ov m}{S\ov\sqrt2}}\right\}a_{16}\left.\right]~;
\label{SOh1}   \\
\chi_{2}&=&\CM_{++}^0~=~-\CM_{--}^0 \nonumber \\
=\widehat\CI\left[\right.
&-&{k\ov M}\left\{a_1-{\gve\ov m}\left({\gve\ov m}\left(a_3-2ma_6\right)
+\gve_\pi a_5+{\eta_0\ov m}a_{11}\right)\right\}
\nonumber \\
&-&{p\ov m}\left\{{\gve_d\ov M}\left({\gve\ov m}a_2+\gve_\pi a_4
+{\eta_0\ov m}a_{10}\right)-{2\gve\gve_d-M^2\ov M}a_6\right\}C
\nonumber \\
&-&{(e_0\eta)\ov m}\left\{{\gve\ov m}a_7+{\eta_0\over m}a_8+\gve_\pi a_{12}
\right\}+2\left\{{p\ov m}\left({\eta_0\gve_d\ov mM}C+{\eta_3k\ov mM}\right)
-i{\eta_2\gve\gve_d\ov m^2M}S\right\}a_9
\nonumber \\
&-&\left\{\eta_0{\gve\gve_\pi-pkC\over m^2}-{k\gve\ov m^2}
\left(\eta_1-i\eta_2\right)S+\eta_3{p\gve_\pi-k\gve C\ov m^2}\right\}
\left\{{p\gve_d\ov mM}a_{13}C-{\gve k\ov mM}a_{14}
+{(e_0\eta)\ov m}a_{15}\right\}
\nonumber \\
&+&2{\gve_d-\gve\ov m}(e_0\eta)a_{16}-
\left\{i\eta_2{2\gve\gve_d-M^2\ov mM}S
+{M\ov m}\left(\eta_1S+\eta_3C\right)\right\}a_{16}\left.\right]~;
\label{SOh2} \\
\chi_{^4_6}&=&\CM_{+-}^\pm~=~-\CM_{-+}^\mp \nonumber \\
=\widehat\CI\left[\right.
&\sqrt2&\left(\begin{array}{c}Cos^2\gvt/2\\Sin^2\gvt/2\end{array}\right)
\left\{\left({p\ov m}a_1\pm ka_6\right)\pm2{p^2k\ov m^2}
\left(\begin{array}{c}Sin^2\gvt/2\\Cos^2\gvt/2\end{array}\right)a_4\right\}
\nonumber \\
&-&{p\ov m}{(e_\pm\eta)\ov m}\left\{{\eta_1-i\eta_2\ov m}a_8+ka_{12}S\right\}
\nonumber \\
&+&2\sqrt2{\eta_3\ov m}\left(\begin{array}{c}Cos^2\gvt/2\\Sin^2\gvt/2
\end{array}\right)a_9\pm{\eta_1-i\eta_2\ov m}\left(2a_9+{p^2\ov m^2}a_{10}
\right){S\ov\sqrt2}
\nonumber \\
&+&{k\ov m}\left\{\left(\eta_1-i\eta_2\right)C-\eta_3S\right\}
\left\{\pm{p\ov m}a_{13}{S\ov\sqrt2}-{(e_\pm\eta)\ov m}a_{15}\right\}
\nonumber \\
&-&2{pk\ov m^2}(e_\pm\eta)a_{16}S\pm{\eta_1-i\eta_2\ov m}
\left(\gve{\gve_\pi\ov m}\mp p{k\ov m}\right){S\ov\sqrt2}a_{16}
\nonumber \\
&+&\sqrt2\left(\begin{array}{c}Cos^2\gvt/2\\Sin^2\gvt/2\end{array}\right)
\left\{{\gve\ov m}\left(\eta_3{\gve_\pi\ov m}\mp\eta_0{k\ov m}\right)
+{p\ov m}\left(\eta_0{\gve_\pi\ov m}\mp\eta_3{k\ov m}\right)\right\}a_{16}
\left.\right]~;
\label{SOh3} \\
\chi_{5}&=&\CM_{+-}^0~=~\CM_{-+}^0 \nonumber \\
=\widehat\CI\left[\right.
&&{p\ov m}\left\{{\gve_d\ov M}\left(a_1-{pk\ov m}a_4C\right)+
{\gve k^2\ov Mm}a_5\right\}S-{p\ov m}{(e_0\eta)\ov m}
\left\{{\eta_1-i\eta_2\ov m}a_8+ka_{12}S\right\}
\nonumber \\
&+&2{\eta_3\gve_d\ov mM}a_9S-{\eta_1-i\eta_2\ov m}\left\{{\gve_d\ov M}
\left(2a_9+{p^2\ov m^2}a_{10}\right)C-{\gve pk\ov m^2M}a_{11}\right\}
\nonumber \\
&-&{k\ov m}\left\{\left(\eta_1-i\eta_2\right)C-\eta_3S\right\}
\left\{{p\gve_d\ov mM}a_{13}C-{\gve k\ov mM}a_{14}
+{(e_0\eta)\ov m}a_{15}\right\}
-2{pk\ov m^2}(e_0\eta)a_{16}S
\nonumber \\
&+&\left\{{2\gve\gve_d-M^2\ov mM}\left(\eta_0{p\ov m}S-{\gve\ov m}
\left\{\left(\eta_1-i\eta_2\right)C-\eta_3S\right\}\right)
+2{\gve pk\ov m^2M}\left(\eta_1-i\eta_2\right)\right\}a_{16}\left.\right]~.
\label{SOh4}
\ee
}
Here
\be
(e_\pm\eta)=\pm{1\ov\sqrt2}\left(\eta_1C\pm i\eta_2-\eta_3 S\right);~
(e_0\eta)=\eta_0{k\ov M}+{\gve_d\ov M}\left(\eta_1S+\eta_3C\right).
\label{AIII6}
\ee
In the spectator case, the integro-operator takes the form (\ref{SO6}).
The calculation is carried out numerically as described in the text.



\newpage

\begin{figure}[h]
\centering
~\\[-1cm]
\epsfig{file=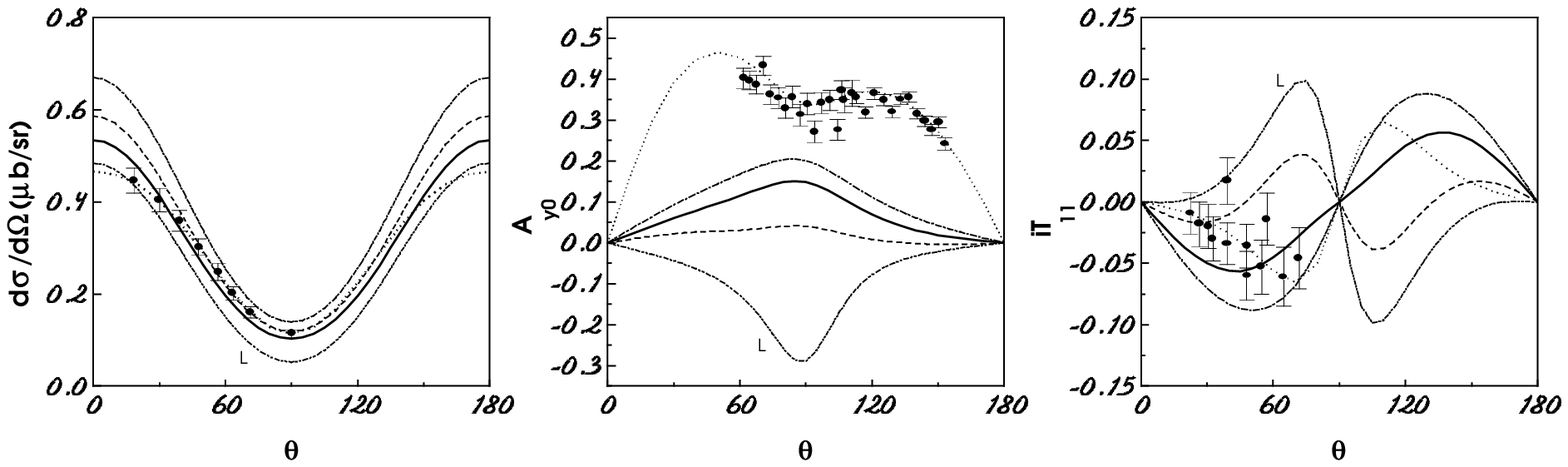,width=14.5cm}
~\\[1cm]
\caption{
Differential cross section $d\gs/d\gO$, asymmetry $A_{y0}$ and vector 
polarization $iT_{11}$ for $pp\to d\pi^+$ as a function of scattering angle 
in the c.m.s. at $T_p=578 MeV$ when the cut-off parameter $\gL$ and mixing 
one $\gl$ varied simultaneously both in the deuteron wave function and in
the $\pi NN$ vertex. The dashed ($\gl=0.6; \gL=1$), solid ($\gl=0.8;
\gL=0.6$) and dot-dashed ($\gl=1;\gL=0.6$) lines correspond to the Gross 
$\CW\CF\CD$ \protect\cite{gross}. The dot-dot-dashed line corresponds to the 
results with Locher's $\CW\CF\CD$ \protect\cite{loch3} ($\gl=1; \gL=1$). 
The dots represent the partial-wave analysis by R. A. Arndt et al. 
\protect\cite{arndt}. The data are from \protect\cite{loch1,loch2,arndt}. 
All spin observables are in the Madison convention.
}
\end{figure} 
\begin{figure}[h]
\centering
~\\[-2cm]
\epsfig{file=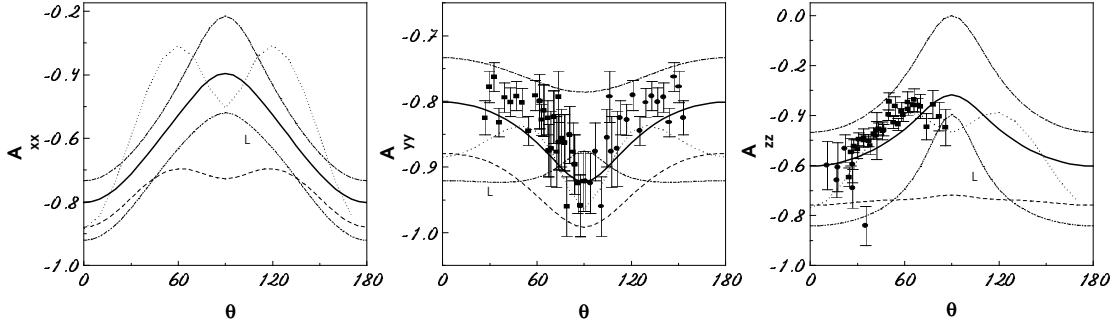,width=14.5cm}
~\\[1cm]
\caption{Spin correlations $A_{ii}$. Notation as in Fig.(1).}
\end{figure}

\end{document}